\documentclass[prl,twocolumn,preprintnumbers]{revtex4-1}
\pdfoutput=1

\usepackage{graphicx,amsmath,amssymb}
\usepackage{hyperref}

\begin{document}

\title{On the unitarity and low energy expansion of the Coon amplitude}



\author{Felipe Figueroa, Piotr Tourkine
\vspace{.1cm}
\\ \small{LAPTh, CNRS et Université Savoie Mont-Blanc, 9 Chemin de Bellevue, F-74941 Annecy, France}}

\email{felipe.figueroa@lapth.cnrs.fr,piotr.tourkine@lapth.cnrs.fr}

\begin{abstract}
  The Coon amplitude is a deformation of the Veneziano amplitude with logarithmic Regge trajectories and an accumulation point in the spectrum, which interpolates between string theory and field theory.
  With string theory, it is the only other solution to duality
  constraints explicitly known and it constitutes an important data
  point in the modern S-matrix bootstrap. 
  Yet, its basics properties are essentially unknown. In this paper we
  fill this gap and derive the conditions of positivity and the low
  energy expansion of the amplitude.
  On the positivity side, we discover that the amplitude switches from
  a regime where it is positive in all dimensions to a regime with
  critical dimensions, that connects to the known $d=26,10$ when the
  deformation is removed. En passant, we find that the Veneziano
  amplitude can be extended to massive scalars of masses up to
  $m^2=1/3$, where it has critical dimension $6.3$.
  On the low-energy side, we compute the first few coupligs of the
  theory in terms of $q$-deformed analogues of the standard Riemann
  zeta values of the string expansion. We locate their location in the EFT-hedron, and find agreement with a recent conjecture that theories with accumulation points populate this space. We also discuss their relation to low spin dominance.
\end{abstract}
\maketitle



The Coon amplitude~\cite{Coon:1969yw,Baker:1971vq,Coon:1972qz} is,
together with the Veneziano amplitude, the only explicitly known
four-point tree-level amplitude that describes an infinite exchange of
higher-spin resonances which solves the duality constraints. It was
discovered as a deformation of the Veneziano amplitude to
non-linear Regge trajectories. The deformation is given in terms of a
parameter $q$ ($0\leq q\leq 1$), which characterises a family of
amplitudes defined by (in units $\alpha'=1$)
\begin{equation}
\label{eq:coon-def}
A_q(s,t) = (q-1)q^{\frac{\log(\sigma)}{\log(q)}\frac{\log(\tau)}{\log(q)}}\prod_{n=0}^{\infty} \frac{\left(\sigma \tau - q^{n}\right)(1-q^{n+1})}{\left(\sigma- q^{n}\right)\left(\tau- q^{n}\right)}
\end{equation}
with
\begin{equation}
  \label{eq:sigma-def}
  \sigma=1+ (s-m^2)(q-1),\quad \tau=1+ (t-m^2)(q-1)
\end{equation}
where $s,t$ are Mandelstam variables (c.f. appendix). This amplitude describes the
scattering of four identical scalars of mass $m^2$.  At $q=0$, it
reduces to a scalar theory, and at $q=1$ it gives back the Veneziano
model:
\begin{align}
  \lim_{q\to0}A_q(s,t)  &=\frac{1}{s-m^2}+\frac 1{t-m^2}+1 \label{eq:q0limit}\\
  \lim_{q\to1}A_q(s,t)  &=A^{V}(s,t)\!=\!-\frac{\Gamma(-s+m^2)\Gamma(-t+m^2)}{\Gamma(-s-t+2m^2)}\label{eq:vene}
\end{align}

Unlike for the Veneziano model, no worldsheet theory was found for the
Coon amplitude, and to this day, its physical origin remains
mysterious. In addition, and what concerns us in this paper, its basic
properties; unitarity conditions and low-energy expansion, are
essentially unknown.

In more recent times, the Coon amplitude was brought forward in the
bootstrap analysis as an exception to the universality of linear Regge
trajectories in~\cite{Caron-Huot:2016icg}, coming from the existence
of an accumulation point in its spectrum, similar to that of the
hydrogen atom, which allows the theory to evade the theorem
of~\cite{Caron-Huot:2016icg}.  Related bootstrap constraints applied
to the Wilson coefficients of effective field theories (EFTs) coming
from unitarity, crossing and analyticity are known to impose
bounds~\cite{Adams:2006sv} that carve theory
islands~\cite{Caron-Huot:2020cmc,Tolley:2020gtv,Arkani-Hamed:2020blm,Bellazzini:2020cot,Sinha:2020win},
and it appears that they are bigger than what is required to describe
the basic theories of the world around
us~\cite{Bern:2021ppb,Caron-Huot:2021rmr,Caron-Huot:2022ugt,Chiang:2022jep}. Even
more interestingly, \cite{Chiang:2022jep} recently conjectured that
the space of gravitational EFTs is actually populated generically of
theories with an accumulation point.
Since the Coon amplitude has an accumulation point and connects
continuously string theory and field theory, it provides an extremely
interesting testing ground to investigate some aspects of these
questions.

\noindent\textit{The main results of our analysis are as follows.}
Firstly, we map the positivity region of the amplitude in the
$(q,m^2)$-space, see fig.~\ref{fig:results}: for each point $(q,m^2)$
we determine the maximal dimension in which no ghosts are exchanged as
intermediate states. This generalises the known $d=(10)\,26$ critical
dimensions of (super)string theory for $m^2=(0)1$. We discover a surprising regime of
the amplitude where we can prove that it is ghost-free in all
dimensions \footnote{Related statements have been made about some
  tensionless strings in the past, but it is not clear if those
  phenomena are related,
  see~\cite{Gamboa:1989zc,Lizzi:1986nv,Karlhede:1986wb,Schild:1976vq}}. This
goes against standard intuition that in high enough dimensions,
string-like theories eventually cease to be unitary. In the other
regime, assisted by analytical arguments, we determine numerically the
positivity surface, which interpolates from infinite critical
dimensions to the standard critical dimensions of string theory.
Along the way, we also realised that the Coon and \textit{a fortiori}
Veneziano amplitudes can be extended to positive $ m^2\to 1/3$, with
corresponding critical dimension $d\simeq6.3$ for the Veneziano
amplitude~\footnote{This fact about intercepts allowed to be negative
  with bound $-1/3$ was probably known, but we have not been able to
  find a trace of it.}.

Secondly, we compute some low energy couplings of the Coon amplitude
in terms of $q$-polylogarithm values, which generalise the known
zeta-values of the string low-energy expansion. We compute explicitly
the first coefficients $g_2(q),g_3(q),g_4(q)$ and map their location
in the space of couplings, comparing to \cite{Caron-Huot:2020cmc}.
We also comment on the connection to the notion of low-spin
dominance of~\cite{Bern:2021ppb}.

\section{The Coon amplitude}
\label{sec:baker-coon-amplitude}

In this section, we review a few facts about the Coon amplitude. Note
firstly that, as a function of $s,t$ alone, as in \eqref{eq:coon-def},
this amplitudes really describes bi-adjoint color-ordered scalars and
should be thought as being stripped of a color factor, in analogy with
similar case in string theory~\cite{Carrasco:2016ldy}. Later, when we
compare the low energy theory to \cite{Caron-Huot:2020cmc}, we
consider the $(s,t,u)$ symmetrized amplitude, given by summing over
the $(s,t)$, $(t,u)$ and $(s,u)$ channels, which corresponds to
ordinary external scalars.

\begin{figure}
  \centering
  \includegraphics{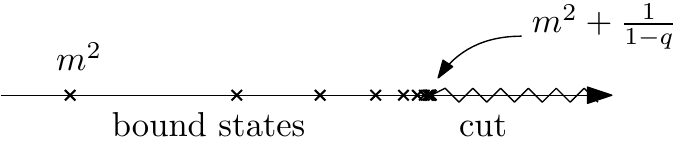}
  \caption{Spectrum of the theory with accumulation point. The poles
    are situated at $s=m^2+\frac{q^n-1}{q-1}$}
  \label{fig:spectrum}
\end{figure}

The spectrum of the amplitude, depicted in fig.~\ref{fig:spectrum},
can be immediately read off eq.~\ref{eq:coon-def}. It has single poles
located at $\sigma=q^n$ which correspond to~\footnote{Note that $\frac{q^{n}-1}{q-1}$ is also known as a $q$-deformed
integer, sometimes denoted $[n]$. When $q\to1$, it reduces to ordinary
numbers, $[n]\to n$ and we recover the spectrum of the Veneziano
model.}
\begin{equation}
  \label{eq:pole-Coon}
  s =m_n^2:= m^2+\frac{q^{n}-1}{q-1},
\end{equation}
These poles build up an accumulation point as $n\to\infty$,
\begin{equation}
  \label{eq:pole-Coon}
  s_* = m^2+\frac{1}{1-q}\,,
\end{equation}
similar to the ionization threshold of the hydrogen atom for instance.
At the accumulation point starts a cut, coming from the
non-meromorphic factor $\sim q^{\log(s)\log(-t)\log(q)}$. As we
explain below, this factor is crucial to ensure polynomial residues,
as was later realised~\cite{Coon:1972qz}. Further analogies with
atomic physics of such an analytic structure were drawn in
\cite{Arik:1973ve} but never made precise~\footnote{Note that the cut
  cannot coincide with a multi-particle threshold, as it would imply
  $4m^2 = m^2 + (1-q)^{-1}$ which can be reached only for $m^2=1/3$
  and $q=0$ which is the trivial theory of
  eq.\eqref{eq:q0limit}}. This makes this amplitude depart from the
strict tree-level, or meromorphic case studied
in~\cite{Caron-Huot:2016icg}. Similar non-analyticities were recently
observed in holographic models for partonic
interactions~\cite{Bianchi:2021sug}, in relation to the bending of
hadron trajectories as $t$ goes to the physical regime and it would be
very interesting to understand this connection further.

On a given pole $s=m_n^2$, the residue of the amplitude is given by a
polynomial of degree $n$ in $t$, corresponding to the exchange of
particles of spins $0$ to $n$ and reads:
\begin{equation}
  \label{eq:residue-Coon}
  \mathrm{Res}(A_q(s,t))\bigg|_{s=m_n^2}= {\prod_{j=0}^{n-1} \left(\frac{\tau-q^{j-n}}{1-q^{-j-1}} \right)}\,,
\end{equation}
On the pole, the non-meromorphic $q$-factor reduces to $\tau^n$
and cancels an inverse factor $\tau^{-n}$ coming from the infinite
product~\footnote{This can be seen by switching to a representation of
  the amplitude where factors of $\sigma$ and $\tau$ appear in the
  denominator of $q^n$, obtained by dividing both the numerator and
  denominator by $\sigma\tau$. In this way, the residue corresponding
  to the infinite product is $\prod(1-q^n/\tau)$ and it is obvious
  that without $\tau^n$ coming from the $q$-factor, this residue would
  not be a polynomial in $t$. Alternatively, one needs to carefully
  treat the infinite products in \eqref{eq:coon-def}, which are not
  convergent by themselves.}. This factor was missed in the original
papers and in the more recent reference
\cite{Fairlie:1994ad}. Interestingly, without it, the amplitude
acquires some severe form of non-locality (non-polynomial residues),
but its unitarity properties do not change much \footnote{We could
  observe this because we started analysing the non-local amplitude
  before realising how it should be fixed}. This illustrates an
interesting aspect of this amplitude, maybe in relation to the
arguments of \cite[appendix C]{Caron-Huot:2016icg} : the presence of
an accumulation point, which in itself did not look dramatic but yet
allowed to escape the axioms of the theorem, seems linked to a subtle
form of non-locality or non-meromorphicity which might explain its
exotic nature.

The Regge and fixed-angle regimes are straightforwardly extracted
\cite{Coon:1969yw,Ridkokasha:2020epy}. In the Regge regime, where
$s\gg-t$ and $t$ fixed, $1/\sigma$ vanishes and the amplitude behaves
as
\begin{equation}
  \label{eq:Regge}
  A_q(s,t)\sim f(t)s^{j(t)},\quad j(t)=\tfrac{ \log((t-m^2)(q-1)+1)}{\log(q)}
\end{equation}
where $f(t) \sim \prod(1-q^n/\tau)$. Since $q<1$, one sees that the
amplitude is suppressed at physical negative $t$ for $m^2\geq0$. In
particular, all bounds on couplings derived assuming standard
twice-subtracted dispersion relations are expected to hold here, as
they use a stronger condition of Regge boundedness
$\frac{A(s,t)}{s^2}\to 0$ at large $s$. We plot $j(t)$ in appendix for
a specific value of the mass, see fig~\ref{fig:plotj}.

In the fixed angle regime, $\frac{s}{-t}$ fixed, both $1/\sigma$ and
$1/\tau$ vanish, therefore we get
\begin{equation}
  \label{eq:Regge}
  A_q(s,t)\sim e^{{\log(s)}{\log(-t) / \log(q)}}\sim s^{\log(s \cos(\theta))/\log(q)}\,.
\end{equation}

\section{Positivity of the amplitude}
\label{sec:unit-accum-point}

We will now present our results on the unitarity of the Coon
amplitude, i.e. the conditions under which no negative norm states are
present as intermediate states. These negative norm states or ghosts
are characterised by negative residues on single poles. More
precisely, near a resonance exchange in the $s$-channel, the amplitude
takes the form
\begin{equation}
  \label{eq:ghosts}
  A(s,t)\sim_{s\to m_n^2} \frac{\mathrm{Res}_n(t)}{s-m_n^2}
\end{equation}

As is standard and reviewed for instance in
\cite{Caron-Huot:2016icg,Arkani-Hamed:2020blm}, Lorentz invariance
implies that $\mathrm{Res}_n(t)$ is a polynomial whose degree
corresponds to the highest spin among the modes of mass $m_n$ being
exchanged, and that it can be further decomposed into angular
eigenfunctions in dimension $d$:
\begin{equation}
  \label{eq:res-expansion}
  \mathrm{Res}_n (t) = \sum_{J=0}^{n} c_{n,J} \mathcal{P}_J^{(d)}(\cos(\theta)).
\end{equation}
Here $\cos(\theta) = 1+\frac{2t}{s-4m^2}$ is the cosine of the
scattering angle and each term in the sum corresponds to a different
particle of mass $m_n$ and spin $L$ being exchanged, and the
$\mathcal{P}_J^{(d)}$ Gegenbauer polynomials (see appendix).

The coefficients $c_{n,J}$ can be obtained using the orthogonality of
the Gegenbauer polynomials, and unitarity demands that these should be
positive, as a negative coefficient implies that the corresponding
exchanged state has negative norm.

In the case of string theory, the no-ghost
theorem~\cite{Goddard:1973qh,Goddard:1972iy} guarantees that such
states decouple from all scattering amplitudes in for $d\leq26$ or $10$. At
the level of the four-point function (the Veneziano amplitude), a recent paper showed that  residues should all be positively
expandable on Gegenbauer polynomials~\cite{Arkani-Hamed:2022gsa} in $d\leq6$, (see also \cite{Maity:2021obe} for
the states on the leading Regge trajectory in $d=4$). It would still be desirable to be able to bridge the gap to $d=10$ or $26$ and maybe the $q\to1$ limit of the Coon amplitude could open an avenue to be combined with the techniques of these papers.

As regards Coon, an early study~\cite{Coon:1974iu} did investigate the
presence of ghosts in the amplitude in four dimensions. Some
partial results were obtained, showing that some regions in the $q,M^2$
parameter space are ghost-free. While their (numerical) method finds ghosts
in four dimensions, we do not, for any values of $q$. This is because we look at a different set-up : for them, the mass of the external
particles $M^2$ was different from $m^2$, the lowest mass of the
amplitude, while for us, $M^2=m^2$.

The more recent reference~\cite{Fairlie:1994ad} which also studied the
problem (which was unaware of~\cite{Coon:1974iu}) is largely
inconclusive and before the present work, nothing was known on
the critical dimensions of the Coon amplitude.

\begin{figure}
  \centering
  \includegraphics[scale=0.45]{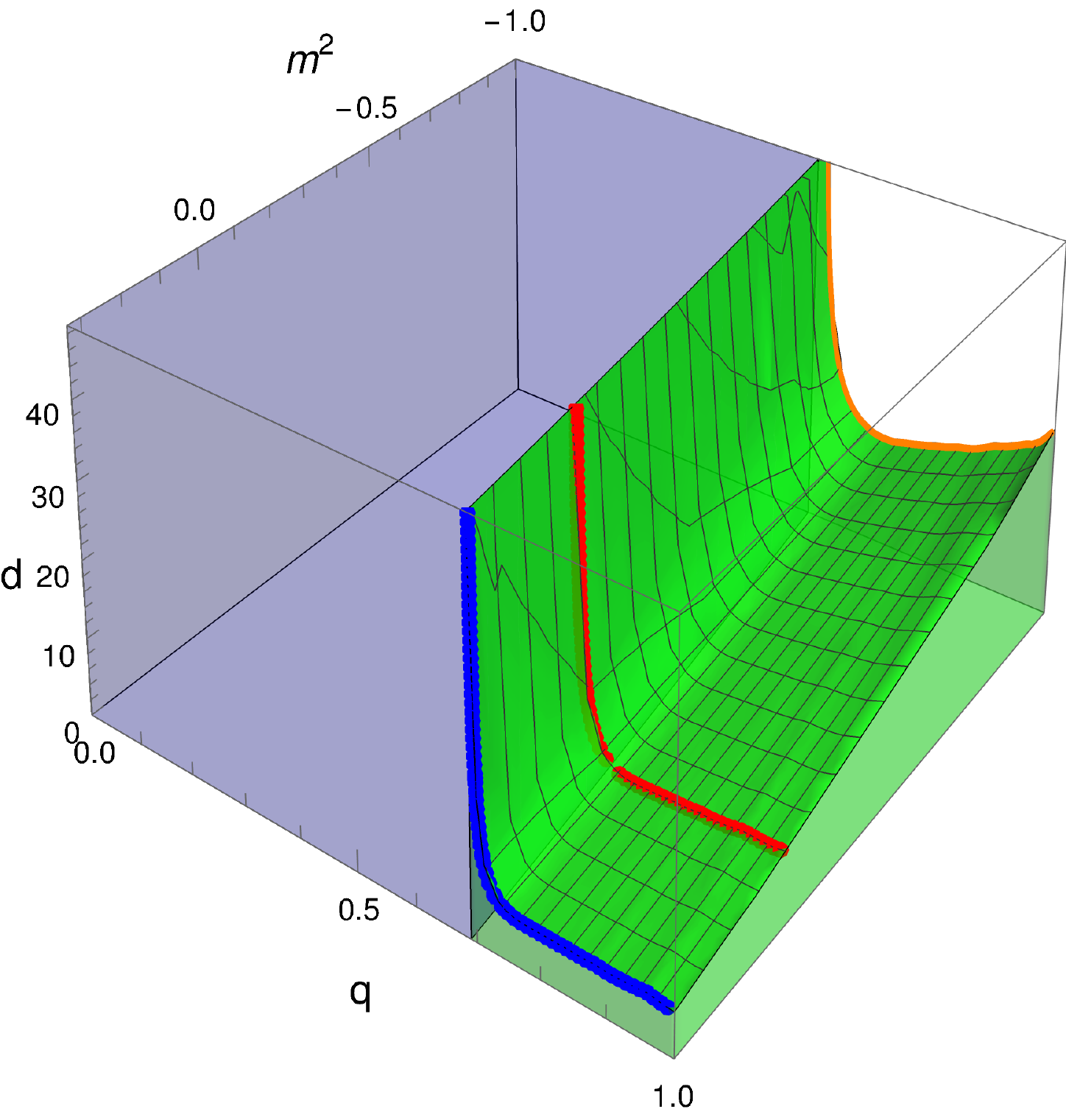}
  \caption{The green surface is the numerically determined boundary of
    unitarity space, everything to the left or bottom of it is
    unitary. The blue shaded region is where we could prove unitarity,
    for all dimensions, and is determined by $q\leq q_c(m^2)$ for all
    $d$. The orange, red and blue curves correspond to $m^2=-1$,
    $m^2=0$ and $m^2=\frac{1}{3}$ }
\label{fig:results}
\end{figure}

\begin{figure}
  \centering
  \includegraphics[scale=0.85]{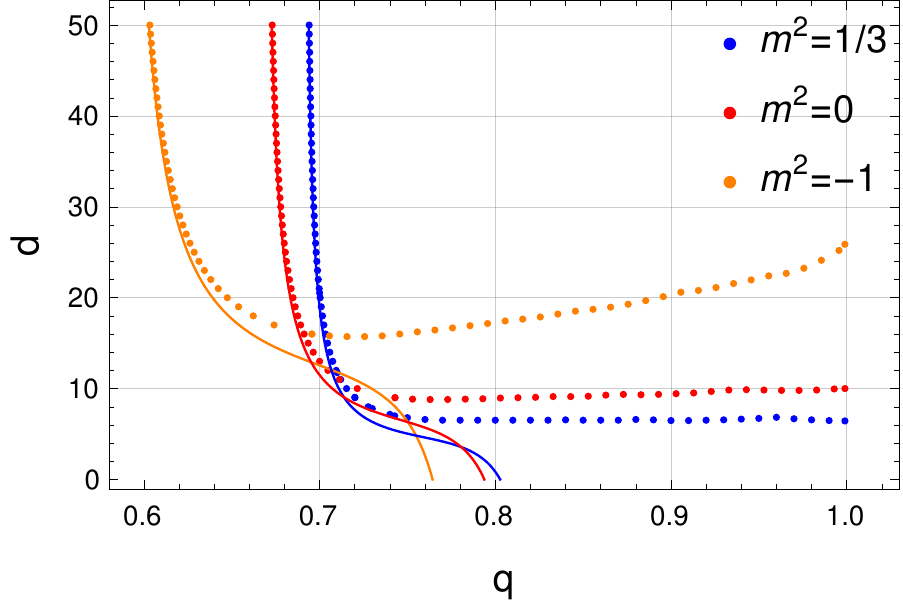}
  \caption{Sections of the surfaced mapped in fig. (\ref{fig:results}) restricted to
    $m^2=-1$, $m^2=0$ and $m^2=\frac{1}{3}$. The dots correspond to numerical results while the solid lines are the large $d$ envelope given by eq. \eqref{eq:d-ratio}. The $q \to 1$ limits of
    the curves match the expected critical dimensions of the Veneziano
    and Neveu-Schwarz amplitudes in the $m^2 = -1$ and $m^2 = 0$
    cases, and suggest the existence of a unitary amplitude with $m^2
    = \frac{1}{3}$ and critical dimension $6.3$.}
\label{fig:results2d}
\end{figure}


Our results are summarized in fig.~\ref{fig:results}. They show
the existence of two regimes for $q$, distinguished critical value
$q\lessgtr q_\infty(m^2)$, where
\begin{equation}
  q_{\infty}(m^2) = \frac{m^2-3+\sqrt{9+2m^2+m^4}}{2m^2}\,.
  \label{eq:qcmsq}
\end{equation}
We find that, for $q<q_\infty(m^2)$ the amplitude is ghost-free in all
dimensions, which we demonstrate analytically. Then, for
$q>q_\infty(m^2)$, are critical dimensions exist and we numerically
determined them, backed by an estimate of the envelope of the critical
dimensions near $q_\infty(m^2)$. Let us now give some details on this
analysis.

\subsubsection{Mass range: $-1\leq m^2\leq1/3$}
Before describing the behaviour in $q$, we describe the range of $m^2$.
For masses $m^2<-1$, the Veneziano amplitude is known to \textit{not}
be unitary (corresponding to intercepts greater than 1). For the Coon
amplitude, a related statement holds : for $m^2<-1$, there is a ghost
at scalar mass-level 2 for $q>\frac2{m^2-1}$. For $q<\frac2{m^2-1}$,
this state becomes a positive-norm state. We gathered solid numerical
evidence that there exists a unitarity range similar to the one we
describe below, for all masses $m^2\to-\infty$. However, to keep a
smooth $q\to1$ limit to string theory we restricted ourselves to
$m^2\geq-1$, but the $m^2<-1$ regime might contain some physics worth
studying~\cite{Masssimo-private}.

The upper bound is imposed on us by unitarity: for
$m^2>\frac{1}{3}$, the Coon amplitude has ghosts in all dimensions
and for all values of $q$. One can check this explicitly by looking
for example at the coefficient $c_{1,1}$, which is given by
\begin{equation}
\label{eq:coef11}
c_{1,1}=\frac{\left(1-3 m^2\right) }{2 (d-3)}q.
\end{equation}
Other coefficients are shown in the appendix in
eq.~\eqref{eq:first-coefs} together with their $q \to 1$
limit. Eq. \eqref{eq:coef11} shows that for $d \geq 4$ this
state is a ghost if $m^2 > \frac{1}{3}$. This stems from the relation
between $\cos \theta$ and $t$, given by
$\cos(\theta) = 1+\frac{2t}{m^2_n-4m^2}$, which becomes singular at
the first resonance for $m^2=\frac{1}{3}$ and causes negative
coefficients beyond this value. 
%
Therefore, we restrict ourselves to the range
\begin{equation}
  -1\leq m^2\leq 1/3\,.
  \label{eq:mass-range}
\end{equation}

Interestingly, this upper bound seems to match the pion intercept fits
of the massive endpoint string models
of~\cite[eq.~4.2]{Sonnenschein:2014jwa}. It would be interesting to
investigate this question further.

\subsubsection{Unitarity in all $d$: $q<q_\infty(m^2)$.}

We shall now prove that the Coon amplitude is unitary in all
dimensions for  $0<q\leq q_{\infty}(m^2)$.
The proof uses the elementary fact that a polynomial with positive
coefficients is positively expandable on Gegenbauer polynomials in all
dimensions, as can be checked on eq.~\eqref{eq:xn-gegen-app}.

The starting point is to write a given residue as a function of
$x=\cos(\theta)= 1+\frac{2t}{m^2_n-4m^2}$,
\begin{equation}
  \label{eq:residue-x}
  \mathrm{Res}A(s,t)\bigg|_{\sigma=q^{n}} = (b_{n})^n
  {\prod_{j=0}^{n-1}(x-x_{j,n})},=
  (b_{n})^n\sum_{k=0}^{n} 
  p_{n,k} x^k.
\end{equation}
where
$ b_n=({1 + 3 m^2 (q-1) - q^n })/{2} $, and
\begin{equation}
x_{j,n} = \frac{3+m^2 (q-1)-2 q^{j-n}-q^n}{1+3 m^2 (q-1)-q^n}
\label{eq:root-xj}
\end{equation}
The residue has all of its coefficients $p_{n,k}$ positive if and only
if all of its roots are positive, thus we want to study the domain in
the $q,m^2$ parameter space where $x_{j,n}\leq0$ for all $j$ and all
$n$. 
Firstly, note that the roots are ordered: since $q<1$, 
if $j<j'$, then $x_{j,n}<x_{j',n}$. Hence, a necessary and sufficient
condition for all the roots to be positive (at fixed $n$) is that
$x_{n-1,n}\leq0$. Since $x_{n-1,n}$ increases monotonically with $n$,
it is further enough to ensure that the limit $n\to\infty$ of
$x_{n-1,n}$ is negative. This amounts to setting $q^n\to0$ in
$x_{n-1,n}$. For this to be negative, we need to have $q$ and $m^2$
related by $q<q_{\infty}(m^2)$ where $q_{\infty}(m^2)$ is defined
above in eq. \eqref{eq:qcmsq}.

We have now concluded that for $q<q_{\infty}$ the residue is a
polynomial with only positive coefficients, which implies that the
amplitude is ghost-free in all dimensions in this region of parameter
space.

\subsubsection{Critical dimensions for $q>q_\infty(m^2)$}

In this range, ghosts are not excluded by our previous argument, and
therefore we performed an extensive numerical study of the sign of the
residues. We computed the Gegenbauer coefficients for the first 50
resonances and all spins for a grid of values of $m^2$ between $-1$
and $\frac{1}{3}$ and varied $q$ and $d$ by small increments to obtain
the critical dimension $d(m^2,q)$ for which all the coefficients
become positive for each $(m^2,q)$, with accuracy $1/25 = 4\%$. This allowed us to map the surface
in parameter space that separates the unitary and non-unitary regions,
given by the green spline in fig.~\ref{fig:results}. In the $q \to 1$
limit, the critical dimensions of the $m^2=-1$ and $m^2=0$ models
match the known values for the Veneziano and Neveu-Schwarz amplitudes
($d=26$ and $d=10$ respectively). Moreover, our results point towards
a possible extremal unitary amplitude with $m^2=\frac{1}{3}$ and
critical dimension $d\simeq6.3$. Those three curves are plotted
specifically in fig.~\ref{fig:results2d}.
We provide some more details on the numerics in the appendix.

One interesting observation from that study is that the scalar ghost
sector seems define the unitarity surface. Below, we give a proof of
this fact in the large $d$ limit and an estimate of the critical
dimensions for $q$ near the critical line which matches the
numerics, see fig.~\ref{fig:results2d}. Surprisingly, while our
arguments fail when $q$ goes closer to $1$, we observe numerically that the scalar ghost criterion continues to hold. Proving this fact fully, maybe using the methods of~\cite{Arkani-Hamed:2022gsa}, would allow to prove exactly the unitarity of Coon (up to finite numerical accuracy in $d$) and of Veneziano amplitude as a function of $m^2$.

The argument about the scalar sector and the critical dimensions goes
as follows. Consider the pole at $\sigma=q^n$.

We want to relate the coefficients $c_{n,L}$ in the Gegenbauer
decomposition of $P_n$ to the coefficients $p_{n,k}$ appearing in
(\ref{eq:residue-x}), which are given in terms of the roots $x_{j,n}$ by
\begin{equation}
p_{n,n-k}=(-1)^k\sum_{i_1 < i_2 <...< i_k} x_{i_1,n}...x_{i_k,n}.
\end{equation}

At fixed $n$, the $c_{n,L}$ coefficient only receives contributions from the
$p_{n,k}$ coefficients with $k \geq L$ and with the same parity. 
Explicitly, for the scalar sector we have
\begin{equation}
  \label{eq:cn0}
  c_{n,0} = p_{n,0}+\frac{p_{n,2}}{-1+d}+\frac{3
    p_{n,4}}{(-1+d)(1+d)}+\dots
\end{equation}

From the form of the roots $x_{j,n}$ one can see that the $p_{L,n}$
coefficients follow a well-defined pattern: when decreasing $q$
starting from values where the amplitude is not unitary, $p_{L,n}$
become positive in decreasing order of
spin. In particular, at some point, all the $p_{n,k}$ coefficients will have
become positive except $p_{n,0}$.

The important point is that right before $p_{n,0}$ turns positive
(when $x_{n-1,n}$ becomes negative), $c_{n,0}$ will become positive by
continuity, as only the first term in \eqref{eq:cn0} remains
negative. Thus, at large $d$, $c_{n,0}$ becomes positive when
\begin{equation}
  \label{eq:d-ratio}
  d= \frac{p_{n,2}}{-p_{n,0}}
\end{equation}
Since $p_{n,0}$ is small and negative, $d$ is large as expected. As
all the other $c_{n,L}$ coefficients do not receive contributions from
$p_{n,0}$, we see that the coefficient $c_{n,0}$ is the last one to
become positive and thus the scalar ghosts are the ones defining the
transition to the unitary regime in the large $d$ limit.

Finally, similar arguments show that this ratio decreases as
$n\to\infty$ and therefore the limit curves enveloppes the region of
unitarity of the amplitude.  It turns out that the limit can be
explicitly evaluated, by summing the infinite $n$ limit of the double
sum
$-p_{n,2}/p_{n,0} = -\sum_{j_1\neq j_2}
\frac1{x_{n,j_1}x_{n,j_2}}$. The resummed function is given in
appendix, and is plotted in \ref{fig:results2d}, and match nicely the
numerical results at large $d$.

\section{Low energy expansion and EFT-hedron}
\label{sec:location-eft-hedron}

Recent times have witnessed a renewal of activity revolving around
implications of dispersion relations, crossing symmetry and
unitarity. Following the ideas of~\cite{Adams:2006sv}, various studies
explored how the Wilson coefficients of {weakly coupled }EFTs are
constrained~\cite{Green:2019tpt,Caron-Huot:2020cmc,Arkani-Hamed:2020blm,Tolley:2020gtv,Bellazzini:2020cot,Sinha:2020win}
and live in some (sometimes small) regions of some positive region
dubbed ``EFThedron''.

In this section, we compute the first few low energy couplings of the
Coon amplitude. Because the Coon amplitude is well behaved at
infinity, and respects analyticity and crossing, it admits dispersion
relations and must fall in those positivity regions. We will see that
the couplings indeed draw 1-dimensional varieties within those
regions, parametrized by the value of $q$.

The first amplitude we consider is an $(s,t,u)$ symmetric version of
Coon, for external \textit{massless} scalars
($m^2=0$) with no color indices:
\begin{equation}
  \label{eq:A-EFT}
  M_q(s,t,u) = A_q(s,t)+A_q(t,u)+A_q(u,s)
\end{equation}
The $(s,t,u)$ symmetry and momentum conservation $s+t+u=0$ allow to
expand at small $s,t,u$ this function in terms of
$\sigma_2=s^2+t^2+u^2$ and $\sigma_3=s t u$, so that
\begin{multline}
  \label{eq:M-taylor}
  M_q(s,t,u) = \frac 1s + \frac 1t + \frac 1u +g_0(q)+ g_2(q) \sigma_2 +\\ g_3(q)
  \sigma_3 + g_4(q) (\sigma_2)^2 + \dots
\end{multline}
where the coefficients of this expansion are classically interpreted
as low energy Wilson coefficients.

A lengthy but straightforward explicit calculation gave us the first
few coefficients, up to $g_4(q)$. Trivially, $g_0(q)=1-q$. The next
ones are given by functions related to $q$-zeta values, for instance
$g_2(q)$ reads
\begin{equation}
  \label{eq:g2q-def}
  g_2(q)=\frac{1}{2} (q-1)^3 \left(3 h_1(q)+5 h_2(q)+2 h_3(q)\right)-\frac{(q-1)^3}{\log (q)}
\end{equation}
where
\begin{equation}
  \label{eq:hm-def}
  h_m(q) = \sum_{n=1}^\infty \frac{q^{nm}}{(1-q^n)^m} := \mathrm{Li}_m(q^m;q)
\end{equation}
can be written in terms of q-deformed polylogarithms as defined for instance in
\cite{schlesinger2001some}, and whose $q$-zeta values are classically
defined as $q$-values of those functions. Note that, compared to
string theory, different orders of $q$-transcendentality appear to be
mixed.
The other couplings $g_3(q)$ and $g_4(q)$ are given in eqs.~\eqref{eq:g3q-def},~\eqref{eq:g4q-def}.  We
also verified that when $q\to1$, they descend to the values given by
the symmetrized sum $A^V(s,t)+A^V(t,u)+A^V(u,s)$:
\begin{equation}
  g_2(1) = -\zeta_3,~g_3(1) = 9/4 ~\zeta_4,~g_4(1)= -\zeta_5/2
\end{equation}

\begin{figure}[t]
  \centering
  \includegraphics[scale=0.7]{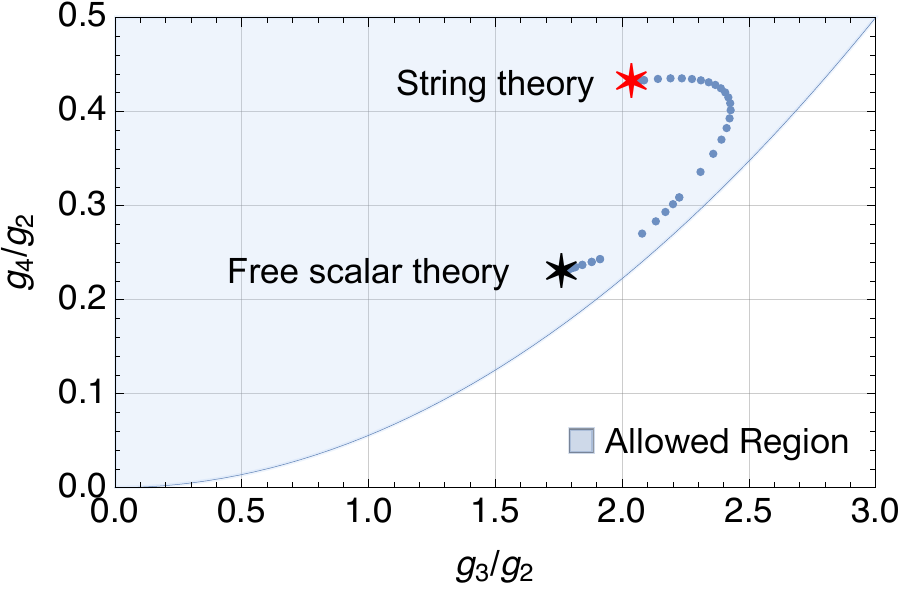}
  \caption{Plot of Wilson coefficients with cut-off $M^2=1$. The scalar limit should be taken with a grain of salt: as $q\to0$, all the couplings $g_{i\geq2}$ are actually sent to zero (see eq.~\ref{eq:q0limit}), but their ratio remains finite. }
  \label{fig:g-plot}
\end{figure}
For generic EFTs, the allowed range of coefficients $g_2,g_3,g_4$ was
determined in \cite{Caron-Huot:2020cmc}, fig.~8, in terms of
dimensionless ratios $\tilde g_3 = g_3 M^2/g_2$ and
$\tilde g_4 = g_4 M^4/g_2$ with $M^2$ given by the scale of the first
massive mode, which in our conventions is $M^2=[1]=1$. We show in
fig.~\ref{fig:g-plot} the value of those ratios. They fall neatly in
the domain determined in~\cite{Caron-Huot:2020cmc}, albeit approaching
tangentially the boundary at intermediate values of $q$.

One can also couple a massless scalar to a \textit{massive} Coon
amplitude, since $0\leq m^2\leq 1/3$ are allowed. These amplitudes
reduce to the extreme scalar case of~\cite{Caron-Huot:2020cmc} of a
coupling the massless scalar to a single massive scalar of mass
$M^2=m^2$ and therefore accumulate to the upper right corner of their
fig.~8. It is not surprising, and the same happens when coupling by a
massless scalar to an amplitude made of massive Veneziano blocks.

\paragraph{Low spin dominance.} The Coon amplitude $A_q(s,t)$,
together with its Veneziano limit, exhibit a form of low-spin
dominance, albeit weaker than that of \cite{Bern:2021ppb}. It is a low spin
dominance were not only the scalar state dominates the partial waves,
but also the spin~1 state~\footnote{We would like to thank Sasha
  Zhiboedov for an illuminating discussion on this point}. Let us
explain how this comes about.

\begin{figure}[t]
  \centering
  \includegraphics[scale=0.35]{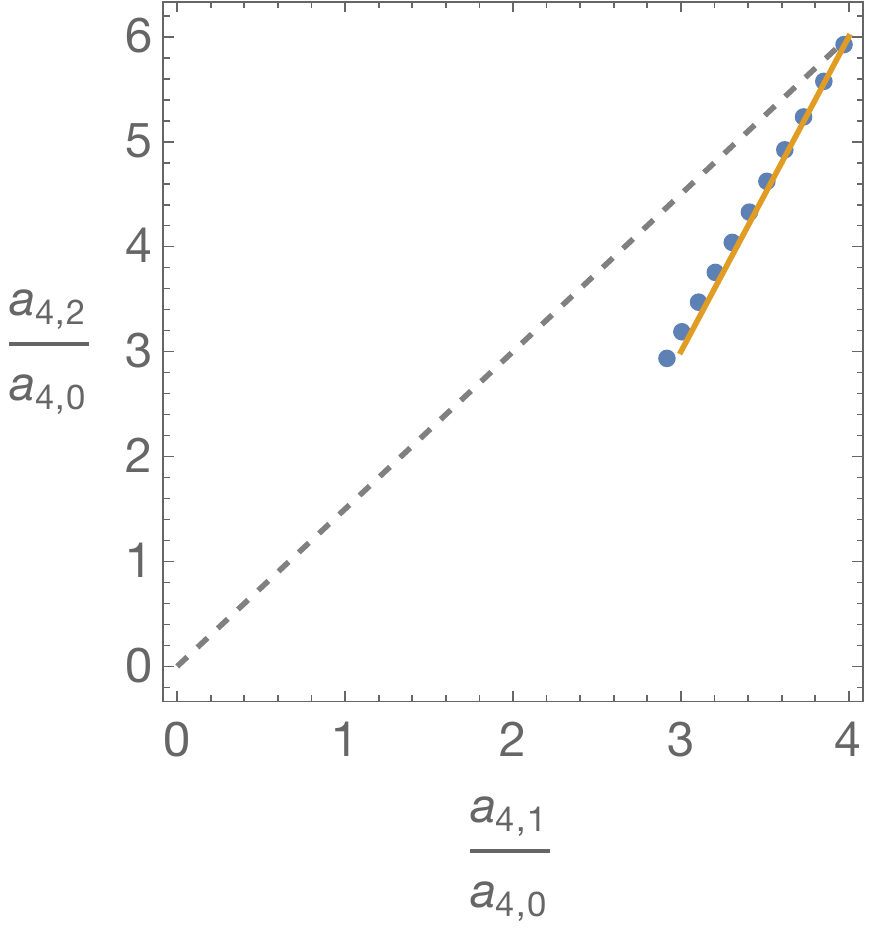}\hspace{24pt}
    \includegraphics[scale=0.45]{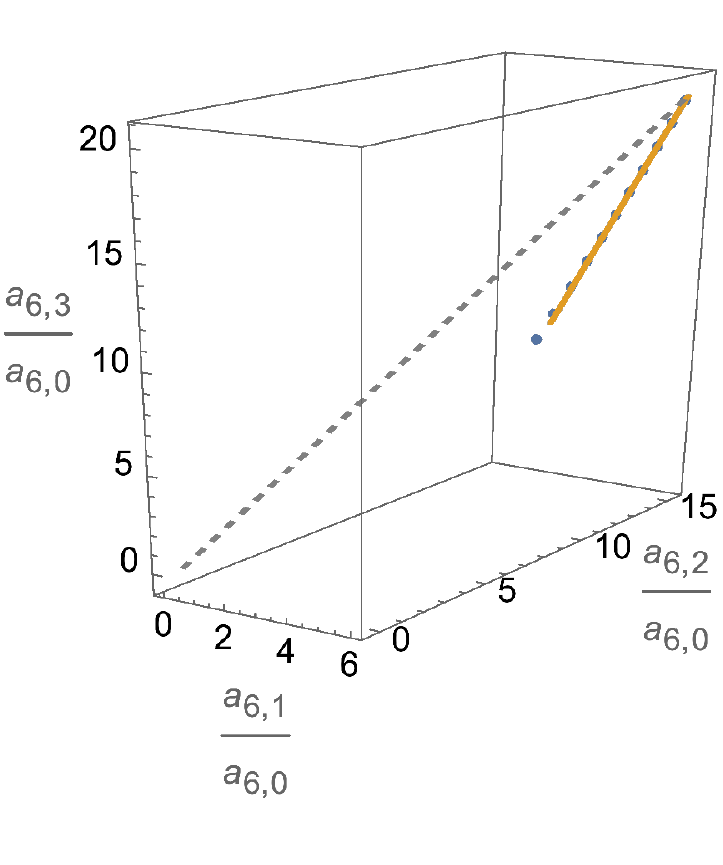}
  \caption{Plots of $a_4$ and $a_6$. Gray dashed : low spin dominance. Yellow : spin 0,1 model. Blue dots : actual points. In
    both cases, Veneziano is in the middle of the graph ($q=1$) and
    the scalar theory is at the intersection with low-spin dominance.}
  \label{fig:lsd}
\end{figure}

Following the conventions of \cite{Bern:2021ppb}, we Taylor expand the
amplitude as
$A_q(t,-s-t) = \frac{1}{s}+\frac{1}{t}+\sum_{p<k} a_{k,p}
s^{k-p}t^p$. In order to match to~\cite{Bern:2021ppb}, we look at the
coefficients at level $k=2,4,6$, which we compare to that of an
amplitude given by a sum of
\begin{equation}
  \label{eq:amplitude-J}
  A^{(J)}(t,u) =(-1)^J \frac{\mathcal{P}_J(1+2s/M^2)}{t-M^2}+\frac{\mathcal{P}_J(1+2t/M^2)}{u-M^2}
\end{equation}
with $J=0,1$. Denoting by $a_{k,q}^{(J)}$ the low energy coefficients
of expanding $A^{(J)}(t,-s-t)$ in powers of $s,t$, the model mentioned
above states that $a_{k,p}\sim a_{k,p}^{(0)}+ q a_{k,p}^{(1)}$. This
corresponds to the straight yellow line in fig.~\ref{fig:lsd}, and can
be seen to match very well the dots near $q=1$ (top-right corner),
where the amplitude matches pure low-spin-dominance, where only $J=0$
contributes.
Within this spin 0-1 model, it is immediate to verify for instance that
\begin{align}
  \label{eq:coefs}
\frac{ a_{2,1}}{a_{2,0}}&=\frac{2 - 2 q}{1 + q}\\
 \frac{a_{4,1}}{a_{4,0}} &=\frac{2 (q+2)}{q+1},~\frac{a_{4,2}}{a_{4,0}} =\frac{6}{q+1}
\end{align}
For $q\in [0,1]$, this implies in particular that $0\leq \frac{a_{2,1}}{a_{2,0}}\leq1$ 
$3\leq \frac{a_{4,1}}{a_{4,0}}\leq4$ and
$3\leq\frac{a_{4,2}}{a_{4,0}}\leq6$. The upper bound correspond to
pure low-spin dominance, while the lower bound is pure spin-$0$ +
spin-$1$ model. While for the coefficients $a_{2,1}/a_{2,0}$, the
bounds are exactly satisfied, at $k=4$ it can be seen that string
theory lies a bit away from that, at
$ (\frac{a_{4,1}}{a_{4,0}},\frac{a_{4,2}}{a_{4,0}})\simeq (2.9,2.9)$.
The relative accuracy of the model is explained by the fact that spin $J$
exchanges come with $q^{J(J+1)/2}$, for which the linear approximation
(spin 0 and 1) is a good approximation away from $q=1$.

\section{Perspectives}
\label{sec:perspectives}

This study opens many perspectives, already mentioned in the text.

Firstly, it resonates very neatly with a conjecture
of~\cite{Chiang:2022jep} that amplitudes with accumulation points
populate the EFT-hedron of gravitational theories away from the small
portion where usual theories seem to live. It would be very important to
study this problem in more details, in relation with the Coon amplitude.

Secondly, it opens a way to attack the question of the positivity of
the Veneziano amplitude recently studied
in~\cite{Arkani-Hamed:2022gsa} thanks to its extra $q$-dependence. In
particular, if one could prove our empirical observation that only the
scalar ghost determines the critical dimension for values of $q$
arbitrarily close to one, maybe using the techniques
of~\cite{Arkani-Hamed:2022gsa}, one could use the smoothness of the
limit to prove positivity of the string theory four-point amplitude.

More generally, it would be of course essential to extend those
results to the $N$-point function. Indeed, to make a statement about
the unitarity of a \textit{theory} (rather than the
\textit{amplitude}), in absence of a no-ghost theorem, one should
indeed prove that ghosts decouple in all exchanges of all
amplitudes. $N$-point functions were proposed as infinite sums in
\cite{Baker:1971vq,GonzalezMestres:1975ord} and factorization was
proven in \cite{Yu:1972fz,Coon:1972te}.

It would also be interesting to study the Coon version of the
Lovelace-Shapiro
amplitude~\cite{Lovelace:1968kjy,Shapiro:1969km}. This amplitude, only
recently understood from string theory~\cite{Bianchi:2020cfc},
constitutes an interesting example where the $N$-point function shows
defaults of unitarity.

Finally, the relation to $q$-zeta values might open an interesting avenue to
produce worldsheet models for the Coon amplitude, and relate to
various integral representation proposed in the
literature~\cite{Chaichian:1992hr,Jenkovszky:1994qg}.

\onecolumngrid

\bigskip

\paragraph{\underline{Acknowledgements}}
We would like to thank Eduardo Casali for collaboration at initial
stages of a related project. We would like to thank Simon Caron-Huot
for some useful discussions, Sasha Zhiboedov for many very stimulating
discussions and comments, Pierre Vanhove for comments on FF's
intership thesis related to this paper, and Massimo Bianchi, Paolo Di
Vecchia and Oliver Schlotterer for useful discussions and detailed
comments on the paper.

\medskip

\twocolumngrid   

\bibliography{biblio.bib}


\section{Appendix}


\appendix

\section{Conventions}
\label{sec:conventions}

\paragraph{Kinematics.}
Here we define our conventions for the article. In the centre-of-mass frame, we have
\begin{eqnarray}
  \label{eq:stu}
  s&=-(p_1+p_2)^2=4E^2=4({\bf p}^2+m^2),\\
  t&=-(p_1+p_4)^2=-2{\bf p}^2(1-\cos(\theta)),\\
  u&=-(p_1+p_4)^2=-2{\bf p}^2(1+\cos(\theta)),
\end{eqnarray}
where $E$ is the center of mass energy, $\bf p$ the momentum transfer
and $\theta$ the scattering angle. Note that the $t\leftrightarrow u$
crossing simply sends $\theta\to \pi-\theta$ in the case of scattering
of identical particles.

The standard relations for the Mandelstam invariants read
\begin{equation}
  \label{eq:momcons}
  s+t+u=4m^2,
\end{equation}

and
\begin{equation}
  \label{eq:scat-ang}
  \cos(\theta) = 1+\frac{2t}{s-4m^2} = \frac{u-t}{u+t}.
\end{equation}

\paragraph{Partial wave expansion.} We follow the conventions of~ \cite{Caron-Huot:2016icg}. The polynomials $\mathcal{P}_L^{(d)}(x)$ diagonalize the Lorentz group Casimir operator in $d$ spacetime dimensions and are defined in terms of hypergeometric functions by
\begin{multline}
\label{eq:gegenbauers-hypergeometric}
\mathcal{P}_{J}^{(d)}(z)={ }_{2} F_{1}\left(-J, J+d-3, \frac{d-2}{2}, \frac{1-z}{2}\right)\\=\frac{\Gamma(1+J) \Gamma(d-3)}{\Gamma(J+d-3)} G_{J}^{\left(\frac{d-3}{2}\right)}(z),
\end{multline}
where $G_{J}^{\left(\frac{d-3}{2}\right)}(z)$ are known as Gegenbauer polynomials. They satisfy the following orthogonality relation:
\begin{equation}
  \label{eq:orth-gegen}
\frac{1}{2} \int_{-1}^{1} d z\left(1-z^{2}\right)^{\frac{d-4}{2}} \mathcal{P}_{J}^{(d)}(z) \mathcal{P}_{\tilde{J}}^{(d)}(z)=\frac{\delta_{J \tilde{J}}}{\mathcal{N}_{d} n_{J}^{(d)}},
\end{equation}
with normalization factors defined by
\begin{equation}
\label{eq:normalization-gegenbauers}
\mathcal{N}_{d}=\frac{(16 \pi)^{\frac{2-d}{2}}}{\Gamma\left(\frac{d-2}{2}\right)}, \quad n_{J}^{(d)}=\frac{(4 \pi)^{\frac{d}{2}}(d+2 J-3) \Gamma(d+J-3)}{\pi \Gamma\left(\frac{d-2}{2}\right) \Gamma(J+1)}.
\end{equation}

Positivity of a given monomial $x^n$ follows explicit expression~
\cite{gradshteyn2007}:
\begin{multline}
  \label{eq:xn-gegen-app}
  \int_{-1}^1 x^{n+2\rho} (1-x^2)^{\frac{d-4}{2}}\mathcal{P}^{(d)}_{n}(x) dx=\\2  \frac{ \Gamma \left(\frac{d-2}{2}\right) \Gamma \left(\rho +\frac{1}{2}\right)\Gamma (n+2 \rho +1)}{2^{n+1} n! \Gamma (1+n) \Gamma (2 \rho +1) \Gamma (n +\rho + \frac{d-1}{2})}.
\end{multline}
for $\rho$ positive integer. When $n$ and $J$ do not have the same
parity modulo 2, $\int_{-1}^1 x^n G_J^{(\nu)}(1-x^2)^{\nu-1/2}$
vanishes. This condition on $\rho$ and $n$, together with a factor of
$2$ are missing in \cite{gradshteyn2007}.

\section{Regge trajectories}
For the interested reader, we show below how the Regge trajectories $j(t)$

\begin{figure}[h!]
    \centering
    \includegraphics[scale=0.7]{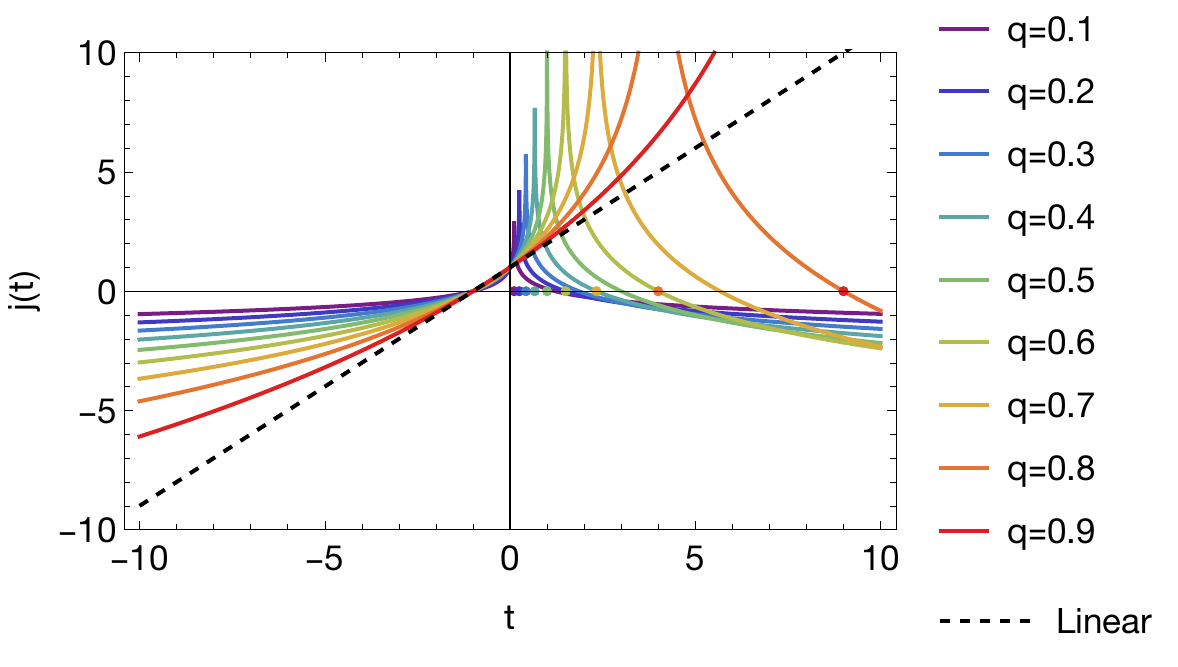}
    \caption{Plot of the real part of $j(t)$ for different values of
      $q$. As $q\to1$, the trajectories unbend and approach the linear
      trajectory of the Veneziano amplitude. Past the accumulation
      point $t=t_*$ (depicted by a dot of the same color as the
      corresponding curve on the $x$-axis), $j(t)$ diverges and then
      acquires a constant imaginary part. Increasing values of the
      mass simply translate the plot to the right.}
    \label{fig:plotj}
\end{figure}

\section{Numerical study}

We mapped the surface that separates the unitary and non-unitary regions in parameter space by a combination of two methods:

To obtain the vertical part corresponding to the high $d$ behaviour we started from a value $q>q_{\infty}(m^2)$ where ghosts are present and
decreased progressively $q$, while keeping $m^2$ and dimension $d$
fixed until all the coefficients became positive. This allowed us to
identify a $d$-dependent critical $q$, which we call $q_c(m^2,d)$, that
determines the boundary of the region of 3-dimensional parameter space
$q,m^2,d$ in which the amplitude is unitary and approaches $q_\infty (m^2)$ in the large $d$ limit.

For the rest of the surface we used the fact that for each $m^2$ there exist a critical dimension below which the
amplitude is ghost-free for all values of $q$. Fixing $q$ and $m^2$ and starting from $d$ lower than this
critical dimension, we
progressively increased $d$ until we found the presence of
ghosts, thus determining the critical surface.

In both cases we computed all the Gegenbauer coefficients for the
first 50 resonances at each value of $q$ with a numerical uncertainty on the critical dimension of $4\%$.

\section{q-values and enveloppe curve}
\label{sec:q-values}

We further find

\begin{align}
  \label{eq:g3q-def}
  g_3(q)=-\frac{3}{2} (q-1)^4 \big(h_1(q){}^2+2 \left(h_2(q)+1\right) h_1(q)+\nonumber\\+h_2(q) \left(h_2(q)+3\right)-h_4(q)\big)+\\+\frac{(q-1)^4 \left(12 \left(h_1(q)+h_2(q)\right)+7\right)}{4 \log (q)}-\frac{3 (q-1)^4}{2 \log ^2(q)}\nonumber
\end{align}
and
\begin{align}
  \label{eq:g4q-def}
\nonumber g_4(q)=  \frac{1}{8} (q-1)^5 \big(h_1(q){}^2+2 \left(h_2(q)+3\right) h_1(q)+\\\nonumber +h_2(q) \left(h_2(q)+21\right)+28 h_3(q)+17 h_4(q)+4 h_5(q)\big)+\\+\frac{(q-1)^5 \left(-12 \left(h_1(q)+h_2(q)\right)-11\right)}{48 \log (q)}+\frac{(q-1)^5}{8 \log ^2(q)}
\end{align}

A related type of computation gave us the enveloppe of the unitarity
curve at large $d$, plotted in fig.~\ref{fig:results2d} which we found to be:

\onecolumngrid
\begin{equation}
\label{eq:ratio-exact}
\lim_{n\to\infty}\frac{p_{n,2}}{-p_{n,0}} =\frac{1}{8}f^2 \left(1+3m^2(q-1) \right)^2 \left(\frac{3}{2}-\log (q) \left(\psi _{\frac{1}{q}}^{(0)}\left(-\frac{\log \left(\frac{q}{f}\right)}{\log \left(\frac{1}{q}\right)}\right)+\log
   \left(\frac{q}{f}\right)+\log (1-q)\right)-\mathrm{Li}_2\left(1;\frac{f}{q}\right)\right),
\end{equation}
\twocolumngrid
with $f=\frac{2}{3+m^2 (q-1)}.$

\subsection{Relation to Veneziano}

Here we review briefly how the Coon amplitude relates
to the Veneziano amplitude by allowing non-linearities and solving an ansatz, following the original reference~\cite{Coon:1969yw}.

The Veneziano amplitude~\cite{Veneziano:1968yb} can be seen as the
unique answer to the following question: Given a theory with spectrum
of resonances $m_n^2$ and assuming that its tree-level 4-point
amplitude can be written in a product representation in terms of its
poles $m^{2}_{n}$ and zeros $\lambda_n$ of the form
\begin{equation}
  \label{eq:prod-rep}
  A(s,t) = \prod_{n=0}^\infty \frac{\alpha_ns+ \beta_nt-\lambda_n
  }{(\alpha'(s-m_n^2))(\alpha'(t-m_n^2))},
\end{equation}
what are the conditions on the $m^{2}_{n}$ and $\lambda_n$ for the
amplitude to have polynomial residues?

It is immediate to realize that with this ansatz, polynomiality
of the residues implies the linear spectrum of the Veneziano amplitude
$m_n^2=4(n-\alpha_0)/\alpha'$ and that the zeroes $\lambda_n$ must be
located at $s+t=m_n^2$ \cite{Coon:1969yw,Baker:1971vq}. 

The Coon amplitude was born out as a way to generalize the Veneziano
amplitude by allowing an extra $s t$ term in the numerator, such that
one is now looking for an amplitude which takes the form
\begin{equation}
  \label{eq:prod-rep}
  A(s,t) = \prod_{n=0}^\infty \frac{\alpha_ns+ \beta_nct + \gamma_n s t -\lambda_n
  }{(s-m_n^2)(t-m_n^2)}
\end{equation}
Demanding that the amplitude reduces to the Veneziano model in some
limit where the deformation parameters go to zero is enough to obtain
the Coon amplitude

In conclusion, the Coon amplitude is the unique solution to the
duality constraints with non-linear trajectories. 
\section{Explicit expression for the coefficients up to level 3}

The Gegenbauer coefficients of the first three levels together with their expansion around $q=1$ are given by:

\onecolumngrid
\begin{equation}
  \label{eq:first-coefs}
\begin{aligned}
c_{0,0}&=1, \qquad c_{1,0}=\frac{\left(m^2-1\right) q}{2} +1 \; = \; \frac{\left(m^2+1\right)}{2} +\frac{\left(m^2-1\right)}{2}  (q-1), \\
 c_{1,1}&= \frac{\left(1-3 m^2\right) }{2 (d-3)}q \; = \; \frac{1-3 m^2}{2 (d-3)}+\frac{\left(1-3 m^2\right) (q-1)}{2 (d-3)}, \\
 c_{2,0} & =\frac{(d-1) \left(q(1-m^2)+q^2-2\right) \left(q \left(q(1-m^2) +q^2-2\right)-2\right)+q^3 \left(-3 m^2+q+1\right)^2}{4 (d-1) (q+1)} \\
 &= \frac{(d-1) \left(m^2+2\right) m^2+\left(2-3 m^2\right)^2}{8 (d-1)}+\frac{(q-1) \left(5 d m^4-10 d m^2-12 d+40 m^4-62 m^2+40\right)}{16 (d-1)}+\mathcal{O}\left((q-1)^2\right), \\
 c_{2,1} & = \frac{q \left(-3 m^2+q+1\right) \left(1-q \left(q(1-m^2)+q^2-2\right)\right)}{2 (d-3) (q+1)} \\
 & =  -\frac{\left(m^2+1\right) \left(3 m^2-2\right)}{4 (d-3)}+\frac{\left(-15 m^4+27 m^2-8\right) (q-1)}{8 (d-3)}+\mathcal{O}\left((q-1)^2\right), \\ 
c_{2,2}&=\frac{q^3 \left(-3 m^2+q+1\right)^2}{2 (d-3) (d-1) (q+1)} \; = \; \frac{\left(2-3 m^2\right)^2}{4 (d-3) (d-1)}+\frac{\left(45 m^4-72 m^2+28\right) (q-1)}{8 (d-3) (d-1)}+\mathcal{O}\left((q-1)^2\right).
\end{aligned}
\end{equation}

\twocolumngrid It can be checked that the limit $q\to1$ reproduces the
coefficients of the Veneziano amplitude and allow to recover the
critical dimensions of (super)string theory for $m^2=-1, 0$. For
instance, for $m^2=-1$ and $q=1$ one has
\begin{equation}
c_{2,0}=-\frac{d-26}{8 (d-1)},
\end{equation}
which shows that the Veneziano amplitude is non-unitary for $d>26$.

\end{document}